\documentclass[english,superscriptaddress,prl,twocolumn,longbibliography]{revtex4-1}
\usepackage[toc,page,title,titletoc,header]{appendix}
\usepackage[T1]{fontenc}
\usepackage{shapepar}
\usepackage{titlesec}
\usepackage{enumerate}
\usepackage{xcolor}
\usepackage{graphicx}
\usepackage[latin1]{inputenc}
\usepackage{epstopdf}
\usepackage{amsthm}
\newtheorem{theorem}{Theorem}
\newtheorem{lemma}{lemma}
\usepackage{amsmath}
\usepackage{shapepar}
\usepackage{color}

\usepackage{amssymb, amsbsy, amsthm}
\makeatletter

\usepackage{babel}
\makeatother

\begin{document}
\title{Lower-Bounding Entanglement in a general Bell scenario }

\author{Liang-Liang Sun }~\email{sun18@ustc.edu.cn}
\affiliation{Department of Modern Physics, University of Science and Technology of China, Hefei, Anhui 230026, China}
\affiliation{Hefei National Research Center for Physical Sciences at the Microscale and School of Physical Sciences, University of Science and Technology of China, Hefei, Anhui 230026, China}

\author{Xiang Zhou}
\affiliation{Hefei National Research Center for Physical Sciences at the Microscale and School of Physical Sciences, University of Science and Technology of China, Hefei, Anhui 230026, China}

\author{Zhen-Peng Xu}
\affiliation{School of Physics and Optoelectronics Engineering, Anhui University, Hefei 230601, People's Republic of China}

\author{Sixia Yu } \email{yusixia@ustc.edu.cn}
\affiliation{Department of Modern Physics, University of Science and Technology of China, Hefei, Anhui 230026, China}
\affiliation{Hefei National Research Center for Physical Sciences at the Microscale and School of Physical Sciences, University of Science and Technology of China, Hefei, Anhui 230026, China}
\affiliation{Hefei National Laboratory, University of Science and Technology of China, Hefei 230088, China}

\date{\today}


\maketitle

\section{Abstract}
  Understanding the quantitative relation between entanglement and Bell nonlocality is a long-standing open problem of fundamental and practical interest.  Here, we tackle this problem in a general Bell scenario.     {We observe that lying in the center of  quantifying these properties are two minimal distances: one from a state to separable states (entanglement), and the other from a correlation to local correlations (nonlocality).} We find that these two distances can be related to each other---the minimal correlation distance provides a lower bound for the minimal state distance,  which  allows us to derive  nontrivial  bounds on  many  entanglement measures with an arbitrary nonlocal correlation.   Moreover,    with the on-hand structural knowledge of entanglement and nonlocality in the  $(n, 2, 2)$ Bell scenario, we refine our estimate significantly. 

\section{Introduction}

 {Quantum entanglement  {as one of the most precious resources in quantum  information science  is}  responsible  for quantum advantages in various information tasks such as quantum teleportation~\cite{PhysRevLett.70.1895}, cryptography~\cite{PhysRevLett.67.661}, quantum algorithms~\cite{PhysRevLett.86.5188, PhysRevA.68.022312, Briegel2009}, and metrology~\cite{Giovannetti2004}. In these applications,   detecting and estimating the  entanglement~\cite{RevModPhys.81.865} carried by a 
 concerned  system  are of central importance.  Conventional approaches~\cite{GUHNE20091}  typically assume that all devices function as intended,  {rendering the whole setup} vulnerable to both malicious interference and unavoidable technical noise.   Alternatively, device-independent   {protocols~\cite{9809039}  (see \cite{2020} for a review) are free from such assumption, enabling} one to  detect and estimate  entanglement  solely with  nonlocal correlations~\cite{Journal,PhysRevLett.121.180503,PhysRevX.7.021042,PhysRevLett.122.060502,PhysRevLett.111.030501}.  {Being  immune to all kinds of noise, }
device-independent  approach  has been extensively applied to entanglement detection~\cite{Journal,PhysRevLett.121.180503,PhysRevX.7.021042,PhysRevLett.122.060502}.   However,  the task of estimating  entanglement in a device-independent manner remains largely unexplored, primarily due to the absence of a  {quantitative link} (precise  relationship) between nonlocality and various entanglement measures. }



 
 {
While entanglement  is  clearly manifested with nonlocal  correlation, the  interplay between  nonlocality  and entanglement   could be very subtle~\cite{PhysRevLett.111.030501, PhysRevLett.109.120402, PhysRevLett.90.080401, PhysRevLett.108.100402, PhysRevLett.115.030404, PhysRevLett.100.090403, PhysRevLett.98.140402, PhysRevA.65.042302, PhysRevLett.74.2619, PhysRevLett.71.1665, Schmid2023understanding}.  { Besides} entangled states not demonstrating Bell's nonlocality~\cite{PhysRevA.40.4277},  entanglement and nonlocality  may not be strictly positively related to each other in some Bell's scenarios~\cite{PhysRevLett.95.210402, Junge2011}.  Till now, the quantitative relation is  mainly established in   minimal Bell scenario, first for  two-qubit systems~\cite{PhysRevLett.89.170401, PhysRevA.88.052105, PhysRevLett.108.100402} then is generalized to arbitrary  dimensional case~\cite{zhangxingjian, PhysRevResearch.7.L022055}.  Such relations highly rely on the computability of  two-qubit entanglement measures~\cite{PhysRevLett.80.2245, Hill_1997,PhysRevLett.80.2245,article}, the minimal entangled state  for a given nonlocality~\cite{Pironio2009}, and  Jordan's lemma~\cite{1570854176056088192, Pironio2009,  PhysRevLett.97.050503}, which are not available in general Bell's scenarios.   Till now,  only negativity has been bounded with  nonlocality~\cite{PhysRevLett.111.030501, PhysRevLett.106.190502}.   Other measures, particularly those with clear operational meanings,  have not yet been quantitatively related  to nonlocality, leaving a significant theoretical gap. }

\begin{figure}
\begin{center}
\includegraphics[width=0.50\textwidth]{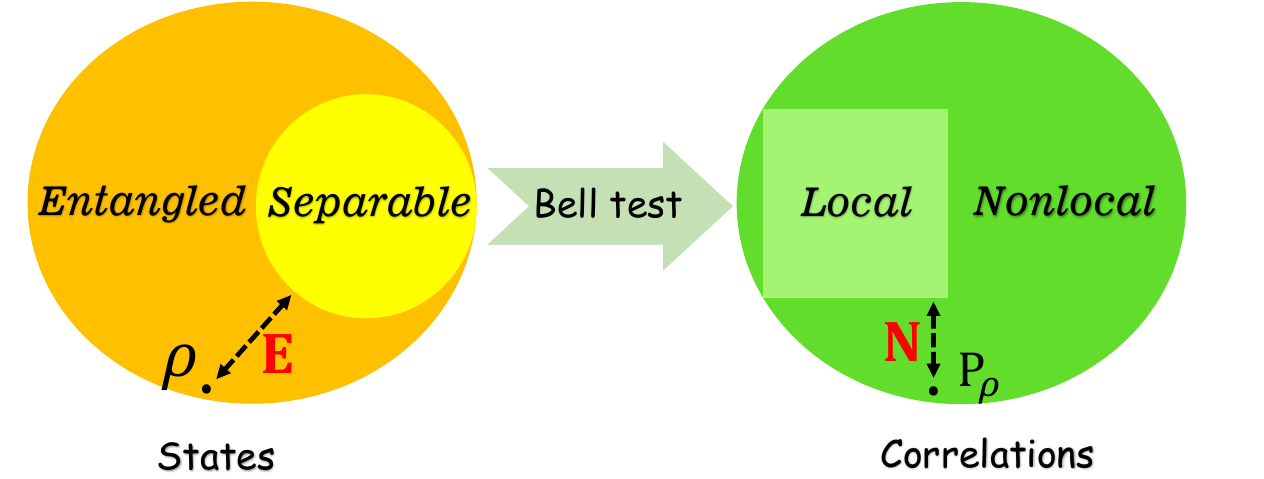}
\end{center}
\caption{ A similarity between the quantification of entanglement and nonlocality: entanglement (E) as the minimal distance between the state of interest and the closest separable state while nonlocality (N) as the minimal distance between the correlation $\mathbf{P}_{\rho}$ and the set of local correlations. }\label{fig2}
\end{figure}


Here we provide a  framework enabling one to estimate various entanglement measures in general Bell's scenarios.  We note that,  while entanglement measures may differ dramatically in mathematical definitions, they can generally be related to  the minimal distance between the state of interest and the set of separable states. Similarly, nonlocality corresponds to the distance from the observed correlation to the closest local correlation.    By this similarity, as shown in Fig. 1, we connect these two distances in two different ways and provide nontrivial lower bounds for  entanglement measures.  As an example,  we  estimate   the extremely weak form of entanglement, namely, bound entanglement, in the device-independent manner.  In some multipartite systems,  our estimation can be asymptotically tight. Furthermore, we demonstrate that when structural knowledge about entangled systems or the Bell scenario is available, e.g., the  $(n, 2, 2)$ Bell scenario, the estimation of entanglement could be improved with  {tightened bound for separable state~\cite{sun2025}}. 

\section{Main Results}
\subsection{Bounding entanglement with nonlocality}
A common approach to quantifying nonclassical properties is to measure their deviation from classical behavior. In this work, we adopt this perspective to analyze the  quantification  of  entanglement and nonlocality and their  relationship.

\emph{A. Entanglement measure via distance---}
Entanglement measures can be defined  (See Supplementary note 1, 2 ) 
as the minimal distance between a concerned state  $\rho$ and the set $\Omega$ of separable states~\cite{PhysRevLett.78.2275, PhysRevA.57.1619}, 
\begin{eqnarray}
E_{\mathcal{D}}(\rho)=\min_{\varrho\in \Omega}\mathcal{D}(\rho, \varrho)\equiv \mathcal{D}(\rho, \Omega),\label{entdis}
\end{eqnarray}
where $\mathcal{D}$ is a distance measure and $\Omega$ denotes set of separable states. Prominent examples are the smallest relative entropy $E_{\rm Re}$,   the  Bures metric  $E_{\rm Bur}$~\cite{PhysRevA.65.062312}, and trace-distance $E_{\rm Tr}$ ~\cite{eisert2006}. There are also other entanglement measures that are not defined in this way,  such as the robust of entanglement $E_{\rm Rob}$~\cite{PhysRevA.59.141} and the measures based on convex-roof construction, e.g.,  the entanglement of formation \cite{PhysRevA.54.3824} $E_{\rm F}$, the concurrence \cite{PhysRevLett.80.2245} $ {E}_{\rm C}$, and the geometric measure of entanglement \cite{PhysRevA.68.042307} $ {E}_{\rm G}$.   They can be lower-bounded by distance-based measures as summarized in the second column of Table 1 (see Supplementary note 2 for the proof).   Consider a  specific state   $\rho=|\phi\rangle\langle \phi|$ with $|\phi\rangle\propto{\sum_{i}|ii\rangle}$. The lower bounds read $E_{\rm Tr}(\phi)\ge1-1/d$   and ${  E}_{\rm F}\geq E_{\rm Re}\geq \log d$ and $  {E}_{\rm C}\geq \sqrt{2(1-{1}/{d})}$,   $  E_{\operatorname{G}(\rho)}\geq 1-1/{d}$, $E_{\rm IF}\geq 1-1/{d} $ and $E_{\rm Rob} \geq d-1$, which all coincide with the exact values of the corresponding measures (see ~\cite{PhysRevResearch.4.023238, PhysRevA.59.141} for $E_{\rm Tr}$ and $E_{\rm Rob}$, and the computation of other measures is straightforward). 

\emph{B.  Nonlocality via distance---}
We now turn to the quantification of nonlocality.  In a general  Bell scenario,  $n$ observers share a joint  $n$-partite  system. Each observer $i$ can randomly choose one local measurement labeled $m_i$ from the   alphabet $\Omega_{i}$ with the outcome specified by  $a_{i}$. The probability distribution of outputting $\vec{a}=(a_{1}, \cdots,  a_{n})$ by performing local measurements $\vec{m}=(m_1, \cdots, m_{n})$ is denoted by $\mathbf{p}_{\vec m}\equiv \{p(\vec{a}|\vec{m})\}_{\vec{a}}$.  The collection $\mathbf{P}=\{ \mathbf{p}_{\vec{m}}\}_{ \vec{m}}$ over all possible measurement settings is called a correlation.  The set  of local correlations specified by $\mathcal L$ is a convex polytope enclosed in some hyper-planes called Bell inequalities~\cite{RevModPhys.86.419}  assuming the form 
\begin{eqnarray*}
\beta(\mathbf{P})\equiv \sum_{\vec{a},\vec{m}}\alpha_{\vec{a}| \vec{m}}\cdot p(\vec{a}|\vec{m})\le \beta({\mathcal{L}})\equiv \max_{\mathbf{P}'\in \mathcal{L}} \beta(\mathbf{P}').
\end{eqnarray*}
where  { $\alpha_{\vec{a}|\vec{m}}$ specifies the coefficient relevant to the probability of $p(\vec{a}|\vec{m})$} and  $\beta(\mathcal{L})$  is the maximum bound of  Bell's inequality  relevant to   local correlations.

As a correlation $\bf P $ is nonlocal if and only if it lies outside $\mathcal{L}$,  we can quantify the nonlocality  of a correlation specified as $N(\bf P)$ by its minimal distance  $\mathfrak{D}({\bf P}, \mathcal{L})$  away from the set of local correlations 
\begin{eqnarray}
  N(\mathbf{P})=\min_{\{\mathbf{q}_{\vec m}\}\in \mathcal{L}}\frac{1}{\tau}\sum_{\vec{m}} D(\mathbf{p}_{\vec{m}}, \mathbf{q}_{\vec{m}})\equiv \mathfrak{D}(\mathbf{P}, \mathcal{L}) , \label{datedis}
\end{eqnarray}
where $\tau=\Pi_{i} t_{i} $ is the total number of choices of  measurement settings $\vec{m}$ with $t_{i}$ specifying  the number of possible measurements in $i$-th side,  and the  classical distance   $D(\mathbf{p}, \mathbf{q})$  is induced by the state distance  $  \mathcal D(\rho,\rho)$ (Supplementary note 1). 
For example, for trace-distance $\mathcal{D}_{\rm Tr}(\rho, \varrho)$ the induced classical distance $D_{\rm Tr}(\mathbf{p},\mathbf{q})$ is the Kolmogorov distance $\sum_{{a}}|p_{a}-q_a|/2$. Here, we refer to   this minimum distance  $\mathfrak{D}(\mathbf{P}, \mathcal{L})$   as distance-based {Bell} nonlocality, and specific cases of trace-distance and relative entropy were defined in Refs~\cite{PhysRevA.97.022111,van2005statistical,zhang2010statistical, PhysRevA.92.010301}. As the local set  $\mathcal{L}$ can be well characterized for any given Bell scenario~\cite{PhysRevLett.85.4418,peres1999all,pitowsky1989quantum,RevModPhys.86.419},
for a given correlation $\mathbf{P}$,   we can in principle compute the distance $\mathfrak{D}(\mathbf{P}, \mathcal{L})$ by an optimization over the set $\mathcal{L}$ even it is hard in general.

Most frequently, nonlocality is verified with the violation of  some specific   Bell inequality as  the amount of violation $\beta(\mathbf{P})-\beta(\mathcal{L})>0$, which can be   normalized  as 
\begin{eqnarray}
\beta_{v}(\mathbf{P})\equiv \frac{\beta\mathbf(\mathbf{P})-\beta(\mathcal{L})}{\beta_{\max}-\beta_{\min}} \label{vob}
\end{eqnarray}
where $\beta_{\rm max}$ and $\beta_{\rm min}$ are the maximal  and minimal quantum bounds for the Bell inequality, and the denominator fixes the arbitrary scaling of  Bell's inequality and the relevant violation.   Henceforth, we may abbreviate $\beta(\mathbf{P})$ as $\beta$ if not introducing  ambiguity.




\emph{Lower bounds of entanglement by  Bell inequalities}   As shown in Fig. 1,  entanglement and nonlocality  can be quantified respectively  with the minimal  state-distance  and the minimal  correlation distance, and in Bell's test states map to correlations  where  separable states always  imply local correlations. By these observations,  we    lower-bound  the minimal state  distance with the minimal local correlation distance as (see method)
\begin{eqnarray}
\mathcal{D}(\rho, \Omega)\geq \mathfrak{D}(\mathbf{P}_{\rho}, \mathcal{L}).~\label{relation}
\end{eqnarray}
In the third column of Table 1, we summaries the lower bounds on entanglement in terms of distance-based nonlocality.  We note that the lower bounds can be tight in some cases, for example,  the maximum violation of  Mermin-Ardehali-Belinskii-Klyshko (MABK) inequality~\cite{PhysRevLett.65.1838, PhysRevLett.67.2761, ABelinski,  PhysRevA.46.5375}  gives the exact value of ${E}_{\rm Tr}$ in the asymptotic limit (with the number of partite tending to infinity).  Our approach can  estimate  extremely weak  bound entanglement (Supplementary note 3). 

We can also  link the normalized Bell violation to entanglement as (see method)
\begin{eqnarray}
\mathcal{D}_{\rm Tr}(\rho, \varrho) \geq \beta_{v}(\mathbf{P}). ~\label{relation2}
\end{eqnarray}
As shown in the fourth column of Table 1, this trace-distance serves as a lower bound for a range of entanglement measures. 
In the following table, we summarize  the  interplay between entanglement and nonlocality.  

\begin{theorem}[Lower-bounding entanglement]\label{theorem}
Entanglement measures (Col.1) have lower bounds in terms of the minimal state distance $\mathcal{D}(\rho, \Omega)$ to the set of separable states (Col.2), distance-based nonlocality ${N(\mathbf{P})}$ of an observed correlation (Col.3), and normalized Bell violation ${ \beta_{v}}$ (Col.4) in a general Bell scenario given in Ref.\cite{PhysRevLett.132.110204}. In an $(n, 2, 2)$ Bell scenario, the latter two lower bounds can be improved to optimized distance-based nonlocality ${\tilde{N}(\mathbf{P})}\ge N(\bf P)$ (Col.5) and optimized violation ${\tilde {\beta}_{v}}\ge \beta_v$ (Col.6).
\begin{table}[h]
$$
\begin{array}{c|ccccc}
\hline\hline
E  & {\mathcal{D}(\rho, \Omega)} &{N}(\mathbf{P}) & {{\beta}_{v}} &\tilde{N}(\mathbf{P})  & { \tilde{\beta}_{v}}\\
\hline
E_{\rm Tr}& \mathcal{D}_{\rm Tr} & N_{\rm Tr}& \beta_{v}& / & /  \\
E_{\rm F},\ E_{\rm  Re}& S & N_{\rm Re}& \frac2{\ln2}\beta^2_{v}& \tilde {N}_{\rm Re}& \frac2{\ln2}\tilde {\beta}^2_{v}\\
E_{\rm C}& \sqrt{2}\mathcal{D}_{\rm Tr} & \sqrt 2 N_{\rm Tr}&\sqrt 2 \beta_{v}&\sqrt 2 \tilde{N}_{\rm Tr}& \sqrt 2 \tilde {\beta}_{v}\\
E_{G}& \mathcal{D}^{2}_{\rm Tr} &N^{2}_{\rm Tr}&\beta^{2}_{v}&\tilde {N}^{2}_{\rm Tr}&\tilde {\beta}^{2}_{v}\\
E_{\rm Bur}& \mathcal{D}_{\rm Bur}&N_{\rm Bur}& {\beta^{2}_{v}} &\tilde {N}_{\rm Bur}&{ \tilde{\beta}^{2}_{v}}\\
E_{\rm Rob}& \frac{\mathcal{D}_{\rm Tr}}{1-\mathcal{D}_{\rm Tr}} &\frac{N_{\rm Tr}}{1-N_{\rm Tr}}&\frac{\beta_{v}}{1-\beta_{v}}&\frac{\tilde{N}_{\rm Tr}}{1-\tilde{N}_{\rm Tr}}&\frac{\tilde{\beta}_{v}}{1-\tilde {\beta}_{v}}\\
\hline\hline
\end{array}
$$
\caption{Main results: five types of lower bounds (Cols.2-6) to various entanglement measures (Col.1).}
\end{table}

\end{theorem}
 We take the entanglement of relative entropy ${\rm E}_{\rm Re}$ as an example to show how these  results are obtained. Following the analysis of Eq.~(\ref{datedis}), we obtain $S(\rho\|\varrho)\geq \sum_{\vec{m}}\frac{1}{\tau}{\rm H}(\mathbf{p}_{\vec{m}, \rho}\|\mathbf{p}_{\vec{m}, \varrho})\geq \min_{\{{\bf q}_{\vec{m}}\}\in \mathcal{L}} \frac{1}{\tau}{\rm H}({\bf p}_{\vec{m}, \rho}\|{\bf q}_{\vec{m}})=N_{\rm Re}$, where relative entropy $S(\rho\|\varrho)\equiv\operatorname{tr}\rho(\log \rho -\log \varrho)$  and $\varrho$ is the closest separable state of $\rho$ with respect to relative entropy.
We also have the relation: $S(\rho\|\varrho) \geq \frac{1}{2\ln 2}({\rm Tr}|\rho-\varrho|)^{2}\geq \min_{\varrho'\in \Omega} \frac{2}{\ln 2}\mathcal{D}^{2}_{\rm Tr}(\rho, \varrho') \geq \frac2{\ln2}{\beta}^2_{v}$,
where  ${\rm Tr}|A|={\rm Tr}(\sqrt{AA^{\dagger}})$  and the first inequality follows from Pinsker's inequality $S(\rho_{1}\|\rho_{2}) \geq  \frac{2}{\ln 2}\mathcal{D}^{2}_{\rm Tr}(\rho_{1}, \varrho_{2}) $.  Using the tightened bound for separable states as shown in the following section, we obtain the improved lower bounds  $\tilde{N}_{\rm Re}$ and $\frac{2}{\ln 2}\tilde{\beta}^2_{v}$. For the other entanglement measures, the entries in the third column follow from Eq.~(\ref{datedis}). The fourth column is derived from the relationship between the relevant distance measure and the trace distance, which is shown Supplementary note 2. The fifth and sixth columns utilize the tightened bound discussed in the next section.

In the framework described above, we establish the interplay between entanglement and nonlocality in a general manner, without relying on their specific structures. While this ensures broad applicability, it can sometimes lead to sub-optimal results.   For instance,   {we consider the  minimal Bell scenario  {where two parties, say, Alice and Bob, share a bipartite state  $\rho$ and  can  randomly  perform binary  measurements  $\hat{A}_{x}$ and $\hat{B}_{y}$, respectively,  with $x, y \in \{0, 1\}$. In this case,} the  Clauser-Horne-Shimony-Holt (CHSH) inequality~\cite{PhysRevLett.23.880} reads:
\begin{eqnarray*}
{\beta}_{\rm CHSH}=A_{0}B_{0}+A_{0}B_{1}+A_{1}B_{0}-A_{1}B_{1} \leq 2,
\end{eqnarray*}
where  $A_{x}B_{y}\equiv {\rm Tr}(\rho \hat{A}_{x}\hat{B}_{y})$ specifies  the expectation value.} For the quantum correlation maximally violating this inequality (the Bell's value is $2\sqrt{2}$), the second column of Table 1 provides a lower bound  
 {$E_{\operatorname{C}}(\rho)\geq 0.15$} for the concurrence while the exact value is $1$ \cite{zhangxingjian}.     By incorporating the  accessible structural knowledge specific to this scenario, our lower bound can be optimized and improved from $0.15$ to $0.71$ (see supplementary note 4).

\section{Tightening the estimation in $(n, 2, 2)$ Bell scenario}

In this section,  we leverage  Jordan's lemma~\cite{1570854176056088192, PhysRevLett.97.050503}  (Supplementary note 5)  {to} improve the lower bounds in the $(n, 2, 2)$ Bell scenario, where there are $n$ parties and each of them has two binary measurements to choose. 
The Jordan lemma  states that,  for two binary projective measurements  $\hat{A}_{0}, \hat{A}_{1}$ acting on a finite-dimensional Hilbert space $ \mathcal{H}$,   there exists a decomposition of $\mathcal{H}=\oplus_{h} \mathcal{H}^{(h)}$ into a direct sum of one or  two-dimensional subspaces $\mathcal{H}^{(h)}$ such that these two measurements are also block diagonalized, i.e.,  $\hat{A}_{0, 1}=\oplus_{h}\hat{A}^{(h)}_{0, 1}$ with $\hat{A}^{(h)}_{0, 1}=\Pi^{(h)}\hat{A}_{0, 1} \Pi^{(h)}$ being projective measurements acting on $\mathcal{H}^{(h)}$, where $\Pi^{(h)}$ specifies the projector onto  the subspace  $\mathcal{H}^{(h)}$.    One can apply Jordan's lemma to each party such that the joint probability has the following decomposition $p(\vec{a}|\vec{m})={\rm Tr}[\rho(\otimes_{i} \hat{A}_{a_i|m_{i}})]={\rm Tr}[\rho \otimes_{i}(\oplus_{(h_{i})} \hat{A}^{(h_{i})}_{a_{i}|m_{i}})]$$=\textstyle {\rm Tr}[\rho\otimes_{i}(\oplus_{(h_{i})} \Pi^{(h_{i})}_{i}\hat{A}^{(h_{i})}_{a_{i}|m_{i}}\Pi^{(h_{i})}_{i})]=\textstyle \sum_{h} {\rm Tr}[\Pi_{h}\rho\Pi_{h}  (\otimes_{i} \hat{A}^{(h_{i})}_{a_{i}|m_{i}})]$
$=\textstyle\sum_{h}r_{h} p_{h}(\vec{a}|\vec{m})$, where  $\hat{A}_{a_i|m_{i}}$ is an operator corresponding to $a_{i}$-th outcome of  measurement  $\hat{A}_{m_{i}}$ and $\Pi_{h}\equiv \otimes_{i} \Pi^{(h_i)}_{i}$ is the projector  onto the $n$-qubit subspace $\otimes_{i}\mathcal{H}^{(h_{i})}_{i}$ with $h=(h_1,\ldots,h_n)$, projecting  $\rho$  to   $\rho_{h}\equiv (\Pi_{h}\rho\Pi_{h})/r_{h}$ with $r_{h}\equiv {\rm Tr}(\rho \Pi_{h})$, and  $\mathbf{P}_{h}=\{p_{h}(\vec{a}|\vec{m})\}_{\vec{a}, \vec{m}}$ is generated  using  qubits  $\rho_{h}\equiv (\Pi_{h}\rho\Pi_{h})/r_{h}$  with the relevant Bell value specified by $\beta_{h}$.  
Regarding the entanglement of state $\rho$,  we have  
\begin{eqnarray*}
{E}(\rho) &\geq &\sum_{h}r_{h}E(\rho_{h}). \label{Jordan2}
\end{eqnarray*}
where the inequality is due to the strong  monotonicity of  entanglement measures~\cite{PhysRevA.99.022338} under the physical operation of projection measurements. Except  the trace-distance measure of entanglement, all other measures considered in this paper  assume this property~\cite{PhysRevA.98.052351}.

By the lemma, one  can  decompose the correlation generated using high-dimensional systems into a convex combination of the   ones using  qubit systems.   Then for these qubit 
 correlations,  we can use the strengthened  bound for separable states provided in Ref~\cite{sun2025} to optimize our estimation.



For qubit systems,  if a measurement setting can generate correlation ${\bf P}$ with Bell's value $\beta$, using a general  separable state $\varrho\in \Omega$ it generates correlation having a tighter bound~\cite{sun2025}  $\beta_{\varrho} \leq  f(\beta)\leq  \beta(\mathcal{L})$,  which defines   a subset of local correlation $\mathcal{L}_{\mathbf{P}}$   as  $\{\mathbf{P'} \mid \beta(\mathbf{P'}) \leq f(\beta)\}$.    This refinement allows us to enhance the entanglement bound for qubit systems from $E(\rho_{h}) \propto \mathfrak{D}_{\mathcal{D}}(\mathbf{P}_{h}, \mathcal{L})$ to $E(\rho_{h}) \propto \mathfrak{D}_{\mathcal{D}}(\mathbf{P}_{h}, \mathcal{L}_{\mathbf{P}_{h}})$. To optimize the bound in terms of  the distance-based nonlocality Eq.(\ref{relation}),  we have 
\begin{eqnarray*}
{E}(\rho) &\geq &\sum_{h}r_{h}E(\rho_{h})  \propto \sum_{h} r_{h} \mathcal{D} (\rho_{h}, \varrho_{h})   \nonumber \\
&\geq  &  \sum _{h} r_{h}\mathfrak{D}_{\mathcal{D}}(\mathbf{P}_{h}, \mathcal{L}_{{\bf P}_h})= \mathfrak{D}_{\mathcal{D}}(\sum _{h} r_{h}\mathbf{P}_{h}, \sum_{h} r_{h}\mathcal{L}_{{\bf P}_h}) \nonumber \\
&=&\mathfrak{D}_{\mathcal{D}}(\mathbf{P}, \mathcal{L}'_{\bf P})\equiv \tilde{N}
\end{eqnarray*}
where $\varrho_{h}$ specifies the closest separable state of  $\rho_{h}$, and $\mathcal{L}'_{\bf P}\equiv \{{\bf P'}|{\bf P'}=\sum_{h} r_{h}{\bf P'}_{h}, {\bf P}'_{h}\in \mathcal{L}_{{\bf P}_{h}}\}$,   and the last inequality is due to the joint-convexity of state distance and $\tilde{N}$ is the optimized distance-based nonlocality.     Consider normalized  Bell's nonlocality $\beta_{v}$ defined by  Eq.\ref{relation2}, we have an optimized one 
\begin{eqnarray*}
\sum _{h} r_{h}E(\rho_{h})&\propto&\sum_{h} r_{h}  \frac{\beta({\mathbf P}_{h})-\beta(\mathcal{L}_{{\bf P}_{h}})}{\beta_{\max}-\beta_{\min}}\geq   \frac{\beta(\mathbf P)-\beta(\mathcal{L}'_{\bf P})}{\beta_{\max}-\beta_{\min}}\nonumber \\
&=&\tilde{\beta}_{v}({\bf P}).
\end{eqnarray*}



As the first example, we consider the CHSH inequality.  {By the tightened bound for separable two-qubit states~\cite{sun2025}},  if projective measurements on qubit systems generate a correlation with Bell's value $\beta$, for any separable state $\varrho$ the Bell's value they can generate is bounded as $\beta_{\varrho} \leq f(\beta)$, where    $f(\beta) =  \sqrt{2+2\sqrt{1-({\beta}^{2}/4-1 )^{2}}}$  when $\beta> 2$ and   $f(\beta)=\beta 
 (\mathcal{L})=2$ when $\beta\leq 2$.   As $f(\cdot )$ is concave  and monotonically decreasing,  
we have  $\sum_{h} r_{h}\beta({\bf P}'_{h})\leq \sum_{h} r_{h} f(\beta_{h}) \leq   f(\sum_{h} r_{h}\beta_h) =  f(\beta )$, where $\beta_{h}=\beta({\bf P}_{h})$ and ${\bf {P}}'_{h}\in {\mathcal{L}_{{\bf P}_{h}}}$.  Then we have  $\beta(\mathcal{L}'_{\bf P})\leq  f(\beta)$ and  the optimized Bell's violation as  
\begin{eqnarray*}
\tilde{\beta}_{v}({\bf P}) =\frac{\beta-f(\beta)}{\beta_{\max}-\beta_{\min}}.\nonumber
\end{eqnarray*}
Specifically, when $\beta=2\sqrt{2}$  we have  $f(\beta)=\sqrt{2}$, and the optimized   Bell's nonlocality is reduced   from $\beta_{v}= \frac{2\sqrt{2}-2}{4\sqrt{2}}\approx 0.15$ to  $\tilde {\beta}_{v}= \frac{2\sqrt{2}-\sqrt{2}}{4\sqrt{2}}\approx 0.25$.

As a second example, we consider the  tripartite system and provide  device-independent estimation of genuine quantum entanglement quantified with tri-partite concurrence  introduced ~\cite{PhysRevA.83.062325} as  $E^{(3)}_{\rm C}\ge   \sqrt{2}\beta^{(3)}_{v}, \sqrt{2}\tilde {\beta}^{(3)}_{v} $  (see Supplementary note 7 for details) with the violation of  tripartite Mermin's  inequality~ ~\cite{PhysRevLett.90.080401}
\begin{eqnarray*}
\beta_{\rm M}\equiv (A_{0}B_{1}+A_{1}B_{0})C_{0}+(A_{0}B_{0}-A_{1}B_{1})C_{1}\leq 2, 
\end{eqnarray*}
 where $\beta^{(3)}_{v}$  is  tri-partite nonlocality attributed to genuine three-partite entanglement and  ${\tilde {\beta}}^{(3)}_{v}$  is the optimized  one using the strengthened bound~\cite{sun2025} $\beta_{\rho_{21}}\leq f_{(21)}(\beta )$ (see Supplementary note 7 for the definition of  $f_{(21)}$), where $\rho_{21}$ is bi-partition separable state.   As $f_{(21)}$ is monotonically decreasing and convex,   we have 
\begin{eqnarray*}
\tilde{\beta}^{(3)}_{v}({\bf P}) =\frac{\beta-f_{(21)}(\beta)}{\beta_{\max}-\beta_{\min}}.\nonumber
\end{eqnarray*}  
Similarly,  the total entanglement is  bounded as   $E_{\rm C}\ge   \sqrt{2}\beta_{v}, \sqrt{2}\tilde {\beta}_{v}$.  In the figures, we have drawn these bounds. Moreover, we also consider a specific  state $\rho_{\lambda}$ that can  produce correlation  with the value $\beta$, and its entanglement  ${E}^{(3)}_{\rm C} (\rho_{\lambda})$ serves as an upper-bound of the entanglement that can be estimated from this Bell's value. 



\begin{figure}
\begin{center}
\includegraphics[width=0.235\textwidth]{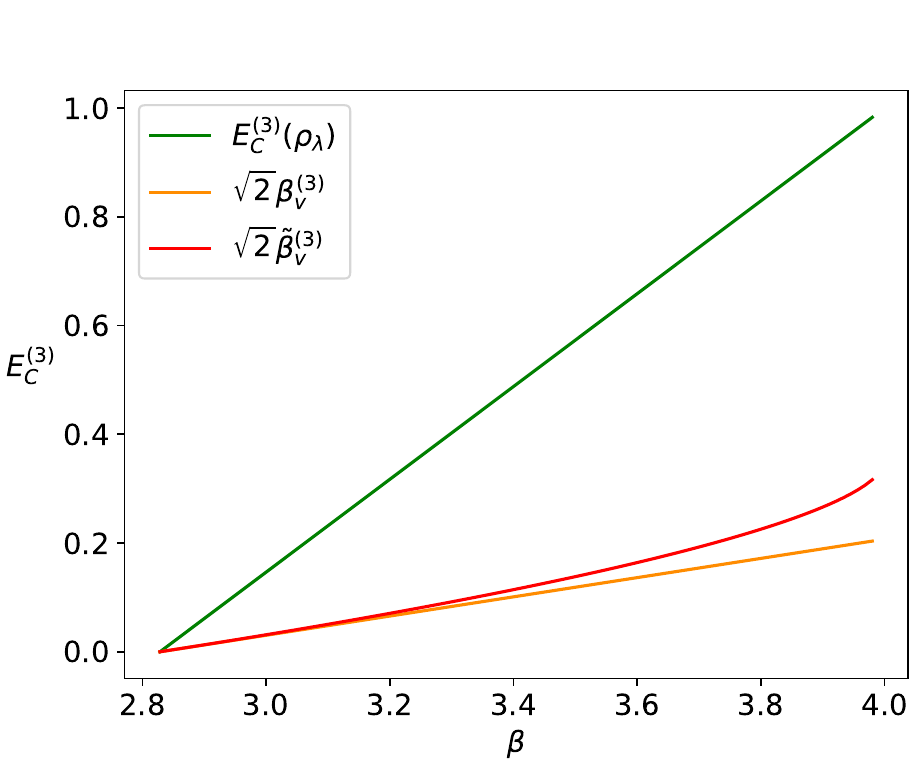}
\includegraphics[width=0.235\textwidth]{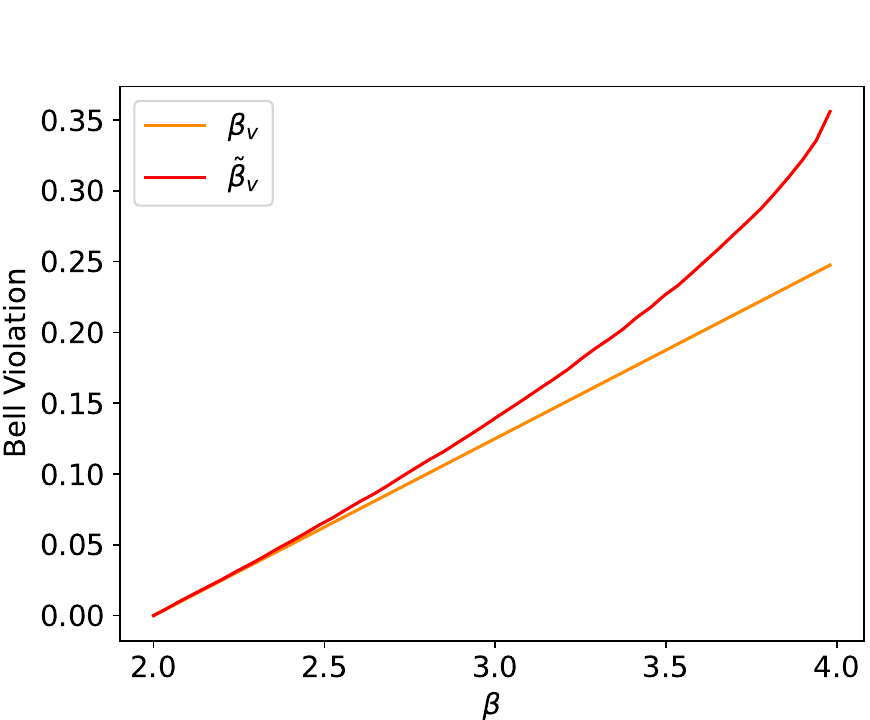}
\end{center}
\caption{ Entanglement estimation based on the {violation of the tripartite Mermin inequality of  a family of state $\rho_{\lambda}$ parameterized by $\lambda$ (See Supplementary note 7).  
On the left: lower bounds on genuine tripartite entanglement $E^{(3)}_{\rm C}$ in terms of normalized Bell violation  $\beta_{v}$ (orange line) and the optimized normalized Bell violation (red curve) $\tilde{\beta}_{v}$, w.r.t. $(21)$ separability, and the entanglement of specific state that can  generate Bell value $\beta$ is given (green line), which is an upper-bound of entanglement (may be not  tight)  that can be verified. On the right: normalized Bell violation before (orange line) and after (red line) optimization w.r.t. full separability.}}\label{fig2}
\end{figure}

\section{Conclusion}\label{sec:conclusion} Here we have explored the quantitative connection between entanglement and Bell nonlocality. Motivated by the observation that these are closely related to two kinds of distance,  namely, the distance of a quantum state from the closest separable state and the distance of correlation from the closest local correlation---we show that these two distances can be generally related to each other, thus providing lower bounds on almost all the important entanglement measures with nonlocal correlation. These lower bounds are nontrivial if and only if the given correlation is nonlocal and can sometimes be asymptotically tight. Noting that these bounds may also be suboptimal at times, we have pinpointed the reasons behind this and shown that the structural knowledge of entanglement and nonlocality plays a key role in optimizing them, for which $(n, 2, 2)$ Bell scenarios are illustrated as examples.

In conclusion, we provide an answer to the long-standing open question in the field of quantum foundations and quantum information science. Our answer not only deepens the understanding of these two fundamental properties and their interplay but also has practical implications, such as device-independent estimation of entanglement. This answer could be optimized if further knowledge about entanglement and nonlocality is obtained.

\section{Methods}
This section provides a bird's-eye view of the proofs for our main results. We refer the reader to Supplementary Note 1 for details.

\subsection{Lower bounding  entanglment to the minimal correlation distance}
Here, we prove Eq.(\ref{relation}), which  shows that   nonlocality as the  correlation distance lower-bounds  the state  distance
\begin{eqnarray}
\mathcal{D}(\rho, \Omega)\geq \mathfrak{D}(\mathbf{P}_{\rho}, \mathcal{L}). \nonumber
\end{eqnarray}
To see this, we consider a  local  measurement   $\vec{m}$ respectively performed on  $\rho$ and its closest separable state  $\varrho \in \Omega$,  {and the relevant statistics are denoted by $\mathbf{p}_{\vec{m}, \rho}=\{{\rm Tr}(\rho\hat{A}_{\vec a|\vec m})\}_{\vec a}$ and $\mathbf{p}_{\vec{m}, \varrho}=\{{\rm Tr}(\varrho\hat{A}_{\vec a|\vec m})\}_{\vec a}$  . It follows from  data processing inequality~\cite{dodonov1999energy, bengtsson2006geometry} that
$
\mathcal{D}(\rho, \varrho)\geq  D\big(\mathbf{p}_{\vec{m}, \rho}, \mathbf{p}_{\vec{m},\varrho}\big),
$  {meaning} that  the state distance is no more than the distance between statistics coming from measurements performed on these two states (Supplementary note 1).}  For a separable state $\varrho\in \Omega$,  the corresponding correlation $\mathbf{P}_{\varrho}=\{\mathbf{p}_{\vec{m}, \varrho}\}_{\vec{m}}\in \mathcal{L}$ is local.   We obtain Eq.(\ref{relation}) by averaging all possible measurement settings $\vec{m}$.

\subsection{Lower-bounding  entanglment with the violation of Bell's inequalities.}
Here we prove Eq.(\ref{relation2}). 
To relate the normalized Bell's violation to entanglement, we use the observation~\cite{PhysRevLett.132.110204}  that
$\mathcal{D}_{\rm Tr}(\rho, \varrho)\geq  |{\rm Tr}[(\rho-\varrho)\mathcal{W}]$, for all normalized operator $-\openone \leq \mathcal{W}\leq  \openone$.  A general operator $\mathcal{W}$ can be normalized as  $\mathcal{W}/|\lambda_{\max}| $, where the $\lambda_{\max}$ is  eigenvalue of $\mathcal{W}$ with the largest magnitude. Although Bell's operator is not characterized a priori in device-independent scenarios, its maximal eigenvalue across all possible measurement settings yields the maximum quantum value. This optimal value can be efficiently computed using semidefinite programming techniques~\cite{PhysRevLett.98.010401}.
 Instead of using $\mathcal{W}/|\lambda_{\max} |$,   we show that Bell's operator $\mathcal{W}$ can be more efficiently  normalized as  $\mathcal{W'}=[2\mathcal{W}-(\beta_{\max}+\beta_{\min})\openone]/(\beta_{\max}-\beta_{\min})$  (Supplementary note 6).  Then we have,   for any $\varrho \in \Omega$ 
\begin{eqnarray}
\mathcal{D}_{\rm Tr}(\rho, \varrho) \geq  |{\rm Tr}[(\rho-\varrho)\mathcal{W}'] \geq 
 \frac{\beta\mathbf(\mathbf{P})-\beta(\mathcal{L})}{\beta_{\max}-\beta_{\min}}\equiv\beta_{v}(\mathbf{P}). \nonumber
\end{eqnarray}
This trace-distance  can lower-bound many other entanglement measures as summarized in the fourth column of Table 1.

\section{DATA AVAILABILITY}
Data sharing is not applicable to this article as no datasets were generated or
analyzed during the current study.

\section{Reference}
\bibliographystyle{apsrev4-2}
\bibliography{entanglement.bib}

\section{Acknowledgements}
\begin{acknowledgments}
L.L.S is supported by development Program of Guangdong Province Grant No. 2020B0303010001.
S.Y. is supported by Innovation Program for Quantum Science and Technology (2021ZD0300804) and Key-Area Research and Development Program of Guangdong Province Grant No. 2020B0303010001.
Z.P.X. acknowledges support from {National Natural Science Foundation of China} (Grant No.\ 12305007),
Anhui Provincial Natural Science Foundation (Grant No.\ 2308085QA29).
\end{acknowledgments}

.

\section{AUTHOR CONTRIBUTIONS}
S. Yu and L.L Sun designed the study and provided the main idea. L.L.Sun, X. Zhou, Z.P. Xu, and S.Yu performed the research, and wrote the manuscript.

\section{ COMPETING INTERESTS} The authors declare no competing interests.

\section{Supplementary note 1. State distance and statistics distance} \label{sect1}

A well-established state distance measure $\mathcal{D}$  assumes some nice properties: 

\begin{itemize}
\item Unitary invariance:  a state distance is invariant under unitary operation, 
$$\mathcal{D}(\rho, \varrho)=\mathcal{D}\big({\rm U}\rho{\rm U}^{\dagger}, {\rm U}\varrho{\rm U}^{\dagger}\big)$$ 
\item  Monotonicity: state distance is  non-increasing  under physical operation specified $\Phi(\cdot )$,  holding data processing inequality $$\mathcal{D}(\rho, \varrho)\geq \mathcal{D}\big(\Phi(\rho), \Phi(\varrho)\big).$$
\item Joint convexity: $$\sum_{i}r_{i}\mathcal{D}(\rho_{i}, \varrho_{i})\geq \mathcal{D}\big(\sum_{i}r_{i}\rho_{i}, \sum_{i}r_{i}\varrho_{i}\big).$$
\end{itemize}

These properties ensure the data processing inequality~\cite{dodonov1999energy, bengtsson2006geometry}  $\mathcal{D}(\rho, \varrho)\geq D(\mathbf{p}_{\rho}, \mathbf{p}_{\varrho})$ with  $\mathbf{p}_{\rho}=\{p_{a|\rho}={\rm Tr}(\rho {\rm M}_{a})\}$, namely,  the state distance is no less than the relevant classical distance $D$ between two distributions from a measurement performed on the states. The reason is as follows. Consider a dephasing operation $\Phi(\rho)\equiv \sum_{a}|a\rangle \langle a|\rho|a\rangle \langle a| =\sum_{a} p_{a|\rho}|a\rangle \langle a|$, where $\{|a\rangle \langle a|\}$ forms an orthogonal basis. By the data processing inequality, $\mathcal{D}(\rho, \varrho)\geq \mathcal{D}[\Phi(\rho), \Phi(\varrho)]$, where the distance on the right-hand side is independent of the choice of orthogonal basis because it is unitarily invariant. Since $\mathcal{D}[\Phi(\rho), \Phi(\varrho)]$ depends only on $\mathbf{p}_{\rho}$ and $\mathbf{p}_{\varrho}$, we can define a classical distance as $D(\mathbf{p}_{\rho}, \mathbf{p}_{\varrho})=\mathcal{D}[\Phi(\rho), \Phi(\varrho)]$. Consequently, the inequality $\mathcal{D}(\rho, \varrho) \geq D(\mathbf{p}_{\rho}, \mathbf{p}_{\varrho})$ follows directly. Unitary invariance ensures that the induced classical distance is uniquely defined.

\section{Supplementary note 2. Entanglement measures are lower-bounded by the minimal state measures}\label{sect2}

One natural method to quantify entanglement is using the minimal distance between the state $\rho$ of interest  and the set $\Omega$ of separable state as~\cite{PhysRevLett.78.2275, PhysRevA.57.1619}, 
\begin{eqnarray}
E_{\mathcal{D}}(\rho)\equiv \mathcal{D}(\rho, \Omega).
\end{eqnarray}

Different choices of state distances may give rise to different entanglement measures~\cite{PhysRevA.56.4452, PhysRevA.57.1619, PhysRevA.65.062312, PhysRevA.73.044301}. For example,   the relative entropy $S(\rho\|\varrho)\equiv\operatorname{tr}\rho(\log \rho -\log \varrho)$  leads to the relative entropy of entanglement $E_{\operatorname{Re}}(\rho)$~\cite{PhysRevA.57.1619}, which  is  an upper bound for the entanglement of distillation~\cite{959270},  and  the trace-distance $\mathcal{D}_{\rm Tr}(\rho_{1}, \rho_{2})\equiv\textstyle 1/{2}\operatorname{\rm Tr}|\rho_{1}-\rho_{2}|$ with $\operatorname{tr}|A|=\operatorname{tr}\sqrt{AA^{\dagger}}$  leads to  the trace-distance measure
$E_{\rm Tr}(\rho)\equiv \min_{\varrho\in \Omega}\mathcal{D}_{\rm Tr}(\rho, \varrho)$ ~\cite{eisert2006}. As  trace-distance has a clear operational meaning as the distinguishability of two known states in experiment~\cite{Helstrom}, the trace-distance entanglement $E_{\rm Tr}$ quantifies the minimal distinguishability of  the state $\rho$ from the set of separable states.   {
By taking the  Bures metric $\mathcal{D}_{\rm Bur}(\rho, \varrho)\equiv 2-2\sqrt{F(\rho, \varrho)}, $
where  ${\rm F}(\rho, \varrho)\equiv [{\rm Tr}(\sqrt{\sqrt{\varrho}\rho\sqrt{\varrho}})]^{2}$ is fidelity,  one has the entanglement   ~\cite{PhysRevA.57.1619} as 
$$E_{\rm Bur}(\rho)\equiv \min_{\varrho\in \Omega}\mathcal{D}_{\rm Bur}(\rho, \varrho).$$
We then have   $ 2-2\sqrt{{\rm F}(\rho, \varrho)} \geq $$ (1+\sqrt{{\rm F}(\rho, \varrho)})({1-\sqrt{{\rm F}(\rho, \varrho)}})= (1-{\rm F}(\rho, \varrho)) \geq  \mathcal{D}^{2}_{\rm Tr}(\rho_{1}, \rho_{2}),$
   where the first inequality is due to  ${\rm F}(\rho, \varrho)\leq 1$ and the last inequality is due to the  Fuchs-van de Graaf (FdG) inequality.  We immediately have $$E_{\rm Bur}(\rho)\geq  \min_{\varrho\in \Omega} ({1}/{2}{\rm Tr}|\rho-\varrho|)^{2}\geq \beta^{2}_{v}$$ where the last inequality is given in  Eq.\ref{relation2}. }

Apart from distance-based entanglement measures, one can define entanglement measures with a convex-roof construction~\cite{Uhlmann1998}. Starting from a measure $E(|\phi\rangle)$ established for an arbitrary pure state $|\phi\rangle $ we extend the measure to mixed states  via convex roof construction as 
\begin{eqnarray}
  {E}(\rho)\equiv \min_{\{|\phi_i\rangle, r_{i}\}} \sum r_{i} E(|\phi_{i}\rangle)
\end{eqnarray}
where the optimization is taken over all the possible pure state ensembles $\rho=\sum_{i}r_{i} |\phi_{i}\rangle \langle \phi_{i}|$. 
The entanglement of formation $E_{\rm F}(\rho)$~\cite{PhysRevA.54.3824},
the concurrence $  E_{\rm C}(\rho)$~\cite{PhysRevLett.80.2245}, and the geometric measure of
entanglement $  E_{\rm G}(\rho)$~\cite{Barnum_2001, PhysRevA.68.042307}  are defined in this way. The entanglement measures for pure states are, respectively,
\begin{eqnarray}
&&{ {E}_{\rm F}}(|\phi\rangle  )\equiv S(\phi_{A}),\\
&&E_{\rm C}(|\phi\rangle)\equiv \sqrt{2-\rm 2Tr(\phi_{A})^{2}},\\
&&E_{\operatorname{G}}(|\phi\rangle)\equiv \min_{|\phi'\rangle \in \Omega}(1-|\langle\phi|\phi'\rangle |^{2}).
\end{eqnarray}
with the reduced density matrix $\phi_{A}={\rm Tr}_{B}(|\phi\rangle \langle \phi|)$ and  $S(\rho)$ denoting the von-Neumann entropy.

{Entanglement of formation}~\cite{PhysRevA.54.3824}   is known to be lower-bounded with relative entropy measure. This is because for a pure state it holds $E_{\operatorname{F}}(|\phi\rangle)=\min_{\varrho\in \Omega}S(\phi\|\varrho)$~\cite{RevModPhys.81.865} and due to the joint convexity of relative entropy~\cite{PhysRevA.57.1619} we have
 \begin{eqnarray}
E_{\rm F}(\rho)&=&\sum_{i}r_{i}S(\phi_{i}\|\varrho_{i})\geq S(\rho\|\varrho)
\geq E_{\rm Re}(\rho),
\end{eqnarray}
 where $\varrho_{i}$ specifies the closest separable state of $|\phi_{i}\rangle $ so that $\varrho\equiv \sum_{i}r_{i}\varrho_{i}\in \Omega$ is separable.

For an arbitrary pure state $|\phi\rangle=\sum_k\alpha_k|a_kb_k\rangle$ in its Schmidt basis, corresponding to which we define a separable state $\varrho=\sum_k|\alpha_k|^2|a_kb_k\rangle\langle a_kb_k|$, then ${\rm Tr} (\phi^{2}_{A})={F}^2(|\phi\rangle,\varrho) $. Let the optimal pure state decomposition for $  E_{\rm C}(\rho)$ be $\{r_{i}, |\phi_{i}\rangle\langle\phi_{i}|\}$, we have lower bound 
 \begin{eqnarray}
  E_{\rm C}(\rho)&= &\sqrt{2}\sum_{i}r_{i}\sqrt{1-{\rm Tr}(\phi_{Ai})^2}\nonumber \\
&=& \sqrt{2}\sum_{i}r_{i}\sqrt{1-F^2(\phi_i,\varrho_{i})}\nonumber \\
&\geq& \sqrt{2}\sum_{i}r_{i}{\mathcal D}_{\rm Tr}(\phi_{i}, \varrho_{i})\nonumber\\ &\geq &\sqrt{2}{\mathcal D}_{\rm Tr}(\rho, \varrho') 
\geq \sqrt{2}E_{\rm Tr}(\rho). \label{councar}
\end{eqnarray}
Here, 
the first inequality follows from the FdG inequality and the second inequality is due to the joint convexity for the trace distance ${\mathcal D}_{\rm Tr} $ with $\varrho'\equiv \sum_{i}r_{i}\varrho_{i}$ being also separable. 

Let $\{r_{i}, |\phi_{i}\rangle\langle\phi_{i}|\}$ be the optimal pure state ensemble achieving the convex-roof for the geometric measure of entanglement $  E_{\rm G}(\rho)$~\cite{Barnum_2001, PhysRevA.68.042307} and denote by $|\phi'_i\rangle$ the product state that maximizes the overlap $|\langle\phi_i|\phi'_i\rangle|^2$. We have lower bound 
       \begin{eqnarray*}
  E_{\operatorname{G}}(\rho)&\equiv& \sum_{i} r_{i}(1-|\langle\phi_i|\phi'_i\rangle|^2)\\
&=&\sum_{i}r_{i}(1-F(\phi_i,\phi'_i)^2 )\\
&\geq& \sum_{i}r_{i}{\mathcal{ D}}^{2}_{\rm Tr}(\phi_{i}, \phi'_{i})  \\
&\ge&\Big(\textstyle\sum_{i}r_{i}{\mathcal{ D}}_{\rm Tr}(\phi_{i}, \phi'_{i})\Big)^{2}\\
&\geq&  {\mathcal{ D}}^{2}_{\rm Tr}(\rho, \varrho')\ge E_{\rm Tr}^2(\rho).
\end{eqnarray*}
Here the first inequality is the FdG inequality, the second inequality follows from the convexity of 
 function $x^2$, and the third inequality is due to the joint convexity of trace distance with  $\varrho'=\sum_{i}r_{i}|\phi'_{i}\rangle\langle\phi'_{i}|$ being separable.

Let $0\leq p\leq 1$ be the smallest positive number such that the mixture $\rho_p=(1-p)\rho+p\varrho$  is separable for  some separable state $\varrho$. 
As $\rho_p$ is separable by definition, we have 
$${E}_{\rm Tr}(\rho)\le \frac12{\rm Tr}|\rho-\rho_p|=\frac p2{\rm Tr}|\rho-\varrho|\le p.$$
We thus arrive at the following lower bound for the robustness of entanglement
 \begin{eqnarray}
{E}_{\rm Rob}(\rho)\equiv \frac{p}{1-p}\geq \frac{{E}_{\rm Tr}(\rho)}{1-{E}_{\rm Tr}(\rho)}.
\end{eqnarray}

\section{Supplementary note 3. Examples for bounding    trace-distance entanglement}\label{sect3}

By the first example, we show that The lower bound in terms of Bell violations for trace-distance measure of entanglement could be asymptotically tight. 
 Let us consider {the $\{n, 2, 2\}$ Bell's scenario}, namely, {$n$ parties, two measurements per party and two outcomes per measurement}. When the total number $n$ of {parties} is odd, the Mermin-Ardehali-Belinskii-Klyshko (MABK) inequality~\cite{PhysRevLett.65.1838, PhysRevLett.67.2761, ABelinski,  PhysRevA.46.5375}  reads
 \begin{equation}\label{M} 
\beta_{\rm MABK}(\mathbf{P}):=\sum_{\stackrel{\vec a,\vec m\in\{0,1\}^n}{\vec 1\cdot\vec m \ {\rm odd}}}(-1)^{\Gamma(\vec m)+\vec 1\cdot\vec a}p(\vec a|\vec m)\stackrel{\mathcal{L}}{\le} 2^{\frac{n-1}2}
 \end{equation} 
 where $\Gamma(\vec m)=\sum_{j>k}m_jm_k$ with $m_{i}\in\{0,1\}$ specifying the setting on {$i$-}th observer and $\vec{1}=(1,1,\cdots, 1)$.
The maximal violation {of} the MABK inequality is achieved by $n$-partite GHZ state and the local  measurements  for each observer can be chosen to be $\{\sigma_x,\sigma_z\}$. In this case, we have $\beta_{\rm max}=-\beta_{\rm min}=2^{n-1}$ \cite{PhysRevA.108.062404} with the classical bound being $\beta(\mathcal L)=2^{\frac{n-1}2}$. As a result, 
$$ 1/{2}=E_{\operatorname{tr}}(\rho_{\rm GHZ})\ge \textstyle \frac12-{2^{-\frac{n+1}2}},$$
where the lower bound
tends to the exact value of trace-distance  entanglement of {$n$-}partite GHZ state in the case of infinite $n$.

By the second example, we  lower-bound bound entanglement. 
  Bound entanglement,  albeit very weak and can not be detected by positive  partial transposition  criterion,  can indeed violate Bell inequality thus still allowing for a device-independent  quantification with our approach. Let us consider a Bell inequality  due to Yu and Oh, in which,   Alice performs $d$ dichotomic measurements labeled with $m_{1}\in \{0, 1, 2,\ldots d-1\}$ while  Bob performs only two measurements with the first one, labeled with 0, being dichotomic and the second one, labeled with 1, having $d$ outcomes. For any local realistic model, it holds~\cite{bound}
\begin{eqnarray}
\beta_{\rm YO}(\mathbf{P})=p(00|01)-\sum_{t=0}^{d-1}p(0t|t0)-\sum_{t=1}^{d-1}p(10|t1)
\stackrel{\mathcal{L} }{\le} 0{.} \nonumber
\end{eqnarray}
It is clear that $\beta_{\rm max}=1$ while $\beta_{\rm min}=-2d+1$ so that a nontrivial lower bound for  $E_{tr}$ {via Bell violation reads}
 \begin{eqnarray}
E_{\operatorname{Tr}}(\rho)\geq \frac{\beta_{\rm YO}(\mathbf{P}_{\rho})}{2d}, \label{ECG}
\end{eqnarray}
which provides a  {device-independent} estimation of entanglement for a family of nonlocal bound entangled states~\cite{bound}.
When $d=3$, the maximum quantum violation of this inequality is  $2.652\times 10^{-4}$, give a lower bound on the corresponding entanglement is  $E_{\rm Tr}(\rho)\geq 4.421\times 10^{-5}$.

\section{Supplementary note 4. Bounding concurrence with different structural knowledge of entanglement in the minimal Bell scenario} \label{sect5}

 {Let us consider the minimal Bell scenario and bound the entanglement measure $E_{\rm C}$ with nonlocality  in terms of the violation of the CHSH inequality by leveraging different levels of structural knowledge:  the computability of entanglement measures for  two-qubit systems;  the properties of the minimal entangled state for a given nonlocality~\cite{Pironio2009}; and  tightened bounds for separable states~\cite{sun2025}. The CHSH parameter for a correlation $\mathbf{P}$ is defined as:
\begin{eqnarray}
\beta_{\rm CHSH}(\mathbf{P})\equiv \sum_{x, y, a, b}(-1)^{xy+a+b}p(ab|xy) \label{CHSH}
\end{eqnarray}
where $x, y \in {0, 1}$ specify the measurement settings and $a, b \in {0, 1}$ the outcomes. }


First, without using any structural knowledge, we have a rather loose lower bound from Eq.(3) in maintext  for the concurrence as follow 
   \begin{eqnarray}
E_{\rm C}&\geq &  \sqrt{2}E_{\rm Tr}\nonumber 
\geq  \sqrt 2 \mathfrak{D}_{\rm Tr}(\mathbf{P}, \mathcal{L}) \nonumber \\
&=&\frac{1}{4\sqrt2} \sum_{a, b, x, y}|p(ab|xy)-p_{l}(ab|xy)| \nonumber \\
&\geq & \frac{1}{4\sqrt2} \left|\textstyle\sum_{a, b, x, y}(-1)^{a+b+xy} \big(p(ab|xy)-p_{l}(ab|xy)\big)\right| \nonumber \\
&\geq & \frac{\beta-2}{4\sqrt 2}.  
\end{eqnarray}
where the first inequality is due to the lower bound of concurrence in terms of trace-distance measure while the second inequality is due to the lower bound of trace distance measure in terms of distance-based nonlocality. In the first equality we have denoted the closest local correlation of $\mathbf{P}$ by $\mathbf{P}_{l}=\{p_{l}(ab|xy)\}$ and we have used $\beta(\mathbf{P}_{l})\leq 2$ in the last inequality. 

Second, by exploiting the knowledge of the maximum and minimal quantum bounds for the CHSH inequality, namely,   $\beta_{\rm max}=2\sqrt{2}$ and $\beta_{\min}=-2\sqrt{2}$, we have the following slightly improved lower bound in terms of  normalized violations  
\begin{eqnarray}
E_{\rm C}\geq \sqrt{2}E_{\rm Tr}(\rho)\geq  \frac{\sqrt{2}[\beta-\beta(\mathcal{L})]}{\beta_{\max}-\beta_{\min}} = \frac{\beta-2}{4}. \label{e2}
\end{eqnarray}

Third, with the knowledge of $(2,2,2)$ Bell scenario in which Jordon's lemma (Supplementary note 5) is applicable, we obtain a further improved lower bound in terms of optimized Bell violation  as
\begin{eqnarray}
E_{\rm C}\geq \sqrt{2}E_{\rm Tr}(\rho)\geq \sqrt{2}\tilde {N}_{v}= \frac{\beta-\sqrt{8-\beta^2}}{8}. \label{e3}
\end{eqnarray} 

Last, in addition to Jordan's lemma we also invoke the minimal entanglement argument. We have the following near optimal lower bound  
\begin{eqnarray}
E_{\rm C}(\rho)&\geq& \sum_{h}r_{h}E_{\rm C}(\rho_{h})\geq \sum_{h}r_{h}E_{\rm C}(\rho_{h|Bell}) \nonumber \\
&\geq &\sqrt{2}\sum_{h}r_{h}E_{\rm Tr}(\rho_{h|Bell})\nonumber\\
& \geq& \frac{1}{2\sqrt{2}}\sum_{h}r_{h}\sqrt{\beta_{h}^{2} -4}\nonumber\\
&\geq& \frac{\beta-2}{4- 2\sqrt{2}}\label{toto}
\end{eqnarray}
The first inequality is due to  strong monotonicity after using Jordan's Lemma, while the second  inequality is due to the minimal entanglement argument, i.e., a two-qubit  state $\rho$ can be transformed into a Bell diagonal state $\rho_{Bell}$ by local operation and classical communications  without changing the relevant Bell value~\cite{Pironio2009}, meaning that it is sufficient for one to consider only Bell diagonal states. 

By denoting Bell state basis $|\psi_{1,2}\rangle=(|00\rangle \pm|11\rangle)/\sqrt{2}$ and  $|\psi_{3,4}\rangle=(|01\rangle \pm|10\rangle)/\sqrt{2} $, a general Bell diagonal state $\rho=\sum_{i}\lambda_{i}|\psi_{i}\rangle \langle \psi_{i}|$ with $\sum_{i}\lambda_{i}=1$  arranged in a  descending order. $\rho$   is entangled when $\lambda_{1}>{1}/{2}$.   Our first observation is that, in terms of trace distance, the closed separable  state $\varrho_{opt}$  to a Bell diagonal state $\rho$  is also a Bell diagonal state. {This is because a state is a Bell diagonal state if and only if it is invariant under unitary operations ${\rm U}_\mu=\sigma_\mu\otimes\sigma_\mu$ for $\mu=x,y,z, 0$} and   ${\rm U}_0=\openone \otimes \openone$.  Let $\varrho_{opt}$ be the closest separable state to $\rho$ in terms of trace-distance and it is clear that
$$\bar\varrho_{opt}=\frac 14\sum_\mu {\rm U}_\mu^\dagger \varrho_{opt}{\rm U}_\mu$$
is also a separable Bell diagonal Bell state. Then we have
\begin{eqnarray*}
{\rm Tr}|\rho-\varrho_{opt}|&=&\frac14\sum_\mu {\rm Tr}|{\rm U}_\mu\rho {\rm U}^{\dagger}_\mu-\varrho_{opt}|\nonumber\\
&=&\frac14\sum_\mu{\rm Tr}|\rho-{{\rm U}_\mu}^{\dagger}\varrho_{opt} {{\rm U}_\mu}|,  \nonumber\\
&\ge &{\rm Tr}\left|\rho-\bar\varrho_{opt}\right|.
\end{eqnarray*}
This means, $\bar{\rho}_{opt}$ is also the closest separable state. Denote $\bar\varrho_{opt}=\sum_{i}\lambda'_{i}|\psi_{i}\rangle \langle \psi_{i}|$  and $\max\{\lambda'_{i}\}\leq 1/2$ as it is separable. We then have 
\begin{eqnarray}
E_{\rm Tr}(\rho_{Bell})=\min_{\lambda'_{i}}\sum_{i}|\lambda_{i}-\lambda'_{i}|=\lambda_{1}-\frac{1}{2}\geq \frac{\sqrt{\beta^{2} -4}}{4},  \nonumber 
\end{eqnarray}
where the minimal is taken over all possible set of normalized $\{\lambda'_{i}\}$ under the constraint $\max\lambda'_{i}\le 1/2$.  In the fourth inequality,   we have used a known result  that Bell value for state $\rho$ is bounded  as $\beta\leq 2\sqrt{1+(2\lambda_{1}-1)^{2}}$
for two-qubit Bell diagonal states~\cite{PhysRevLett.89.170401, PhysRevA.88.052105, PhysRevA.101.042112}.

\section{Supplementary note 5. Jordan's Lemma}\label{sect6}

 {Jordan's Lemma  is found independently in Ref~\cite{1570854176056088192} and \cite{PhysRevLett.97.050503}, which enables one to  bypass the issue of dimension and  work within qubit spaces. }

\begin{lemma}
    For two binary projective measurements  $\hat{A}_{0}, \hat{A}_{1}$ acting on a finite-dimensional Hilbert space $ \mathcal{H}$,   there exists a direct sum of this system decomposition of $\mathcal{H}=\oplus_{h} \mathcal{H}^{(h)}$ into a direct sum of one- or two-dimensional subspaces $\mathcal{H}^{(h)}$ such that these two measurements are also block diagonalized, i.e.,  $\hat{A}_{0, 1}=\oplus_{h}\hat{A}^{(h)}_{0, 1}$ with $\hat{A}^{(h)}_{0, 1}=\Pi^{(h)}\hat{A}_{0, 1} \Pi^{(h)}$ being projective measurements acting on $\mathcal{H}^{(h)}$, where $\Pi^{(h)}$ specifies the projector onto  the subspace  $\mathcal{H}^{(h)}$.   Furthermore, $\hat{A}^{(h)}_{0, 1}$ can be assumed to be  real operators.
\end{lemma}

Concerning a correlation $\mathbf P$ in a $(n, 2, 2)$ Bell scenario with a Bell value $\beta(\mathbf P)$, one can apply Jordan's lemma to  each party such that the joint probability has the following decomposition
\begin{eqnarray*} \nonumber
p(\vec{a}|\vec{m})&=&{\rm Tr}[\rho(\otimes_{i} \hat{A}_{a_i|m_{i}})]\\
&=&{\rm Tr}[\rho \otimes_{i}(\oplus_{(h_{i})} \hat{A}^{(h_{i})}_{a_{i}|m_{i}})]\\
&=&\textstyle {\rm Tr}[\rho\otimes_{i}(\oplus_{(h_{i})} \Pi^{(h_{i})}_{i}\hat{A}^{(h_{i})}_{a_{i}|m_{i}}\Pi^{(h_{i})}_{i})] \nonumber \\
&=&\textstyle \sum_{h} {\rm Tr}[\Pi_{h}\rho\Pi_{h} ( \otimes_{i} \hat{A}^{(h_{i})}_{a_{i}|m_{i}})]\\
&=&\textstyle\sum_{h}r_{h} p_{h}(\vec{a}|\vec{m}),\label{demco}
\end{eqnarray*}
where $\hat{A}_{a_i|m_{i}}$ is operator corresponding to $a_{i}$-th outcome of  measurement  $\hat{A}_{m_{i}}$ and $\Pi_{h}\equiv \otimes_{i} \Pi^{(h_i)}_{i}$ is the projector  onto the $n$-qubit subspace $\otimes_{i}\mathcal{H}^{(h_{i})}_{i}$ with $h=(h_1,\ldots,h_n)$, projecting  $\rho$  to   $\rho_{h}\equiv (\Pi_{h}\rho\Pi_{h})/r_{h}$ with $r_{h}\equiv {\rm Tr}(\rho \Pi_{j})$.   Due to the direct-sum representation,  a Bell correlation $\mathbf{P}$ in the $(n, 2, 2 )$ scenario thus can be seen as a mixture of correlations generated by performing projection measurements on $n$-qubit system  $\mathbf{P}=\sum_{h}r_{h}\mathbf{P}_{h}$, where $\mathbf{P}_{h}=\{p_{h}(\vec{a}|\vec{m})\}_{\vec{a}, \vec{m}}$ is generated from $n$-qubit state $\rho_{h}$  with the relevant Bell value specified by $\beta_{h}$.  Then one has lower bound 
\begin{eqnarray*}
{E}(\rho) &\geq &\sum_{h}r_{h}E(\rho_{h}) \label{Jordan2}
\end{eqnarray*}

In the above, the first inequality is due to the strong  monotonicity of entanglement measure~\cite{PhysRevA.99.022338}   under physical operation of projection measurements. Except the trace-distance measure of entanglement, all other measures considered here  assume this property~\cite{PhysRevA.98.052351}.


\section{Supplementary note 6. Optimization the bound in terms of  normalzied Bell violation}\label{sect7}

Entanglement can be  estimated with  nonlocality-based witness ~\cite{PhysRevLett.132.110204},  
$$\mathcal{W}=\beta(\mathcal{L})\openone -\mathcal{B},$$
where $\mathcal{B}=\sum_{\mathbf{a},  \mathbf{s}}a_{\mathbf{a}, \mathbf{s}} \hat{A}_{a_{1}|s_1}\otimes\cdots \otimes\hat{A}_{a_{n}|s_n}$ is a Bell operator  and and $\beta(\mathcal{L})$   is  the LHV  bound of the Bell quantity $\beta$, which  is also the maximum value of Bell quantity over separable states.

Note that ${\rm Tr }|\rho-\varrho|\geq {\rm Tr }[(\rho-\varrho)\mathcal {W'}]$ for any normalized  Hermitian operator  $-\openone\leq \mathcal{W'}\leq \openone$. There are different ways to normalize a witness operator. For example, 
\[
\mathcal{W}' = \frac{2\mathcal{W} - (\lambda_{+} + \lambda_{-})\openone}{\lambda_{+} - \lambda_{-}},
\]
where $\lambda_{\pm}$ are the largest and smallest eigenvalues of $\mathcal{W}$. Note that $\lambda_{+} \leq \beta(\mathcal{L}) - \beta_{\min} $ and  $\lambda_{\min} \geq \beta(\mathcal{L}) - \beta_{\max}$. Thus, 
\[
\lambda_{+} - \lambda_{-} \leq \beta_{\max} - \beta_{\min},
\]
where $\beta_{\max}$ and $\beta_{\min}$ are the quantum bounds for Bell's inequality as well as the largest and smallest eigenvalues of the Bell operator $\mathcal{B}$ over all possible choices of quantum measurements. These bounds can be computed with a semidefinite program.   
We have 
\begin{eqnarray}
{\rm Tr}|\rho-\varrho|&\geq &{\rm Tr}[(\rho-\varrho)\mathcal{W}']\geq - \textstyle \frac{2{\rm Tr}[(\rho-\varrho)\mathcal{W}]}{\beta_{\max}  -\beta_{\min}} \nonumber \\
&\geq &2\textstyle  \frac{\beta-\beta(\mathcal{L})}{\beta_{\max}  -\beta_{\min}}=2\beta_{v}.
\end{eqnarray}
Immediately, $E_{\rm Tr}(\rho)\geq \beta_{v}$.   
Consider a general   $(n, 2, 2)$ scenario  and Jordan's lemma, using the strengthened bound for separable states we have 
\begin{eqnarray}
E(\rho)&\geq &\sum_{h}r_{h}E(\rho_{h})\propto
\sum_{h} r_{h}\textstyle\frac{\beta(\mathbf{P}_{h})-\beta(\mathcal{L}_{\mathbf{P}_{h}}) }{\beta_{\max}  -\beta_{\min} } \nonumber\\
&\geq &\textstyle \frac{\beta-\beta(\mathcal{L}'_{\mathbf{P}}) }{\beta_{\max}  -\beta_{\min} }
\equiv \tilde {\beta}_{v}. \label{jordan4}
\end{eqnarray}
where the first inequality is  strong  monotonicity of entanglement measure and  $\mathcal{L}'_{\bf P}\equiv \{{\bf P'}|{\bf P'}=\sum_{h} r_{h}{\bf P'}_{h}, {\bf P}'_{h}\in \mathcal{L}_{{\bf P}_{h}}\}$, for which, it is clear  $\beta(\mathcal{L}'_{\bf P})=\sum_{h} r_{h}\beta({\bf P}'_{h})\leq \sum_{h} r_{h} f(\beta({\bf P}_{h})) \leq   f(\sum_{h} r_{h}\beta({\bf P}_h)) =  f(\beta ) $, where we have assumed the concavity and monotonically decreasing of $f$. 

Then entanglement estimation can be optimized as 
\begin{eqnarray}
{E}(\rho) &\geq &\sum_{h}r_{h}E(\rho_{h})  \geq \sum_{h}r_h\mathcal{G}_{v}(\beta_{v_{h}}) \nonumber \\
&\geq & \mathcal{G}_{v}(\sum_{h}r_{h}\beta_{v_{h}})  
=    \mathcal{G}_{v}(\tilde {\beta}_{v}). \label{Jordan3}
\end{eqnarray}
where  the second inequality is due to the bound of entanglement as a function  $\mathcal{G}_{v}$ of  normalized Bell's violation (as shown in the fourth column in the Table in the main text). For instance, when the entanglement measure is  the relative entropy of entanglement, we have $\mathcal{G}_{v}(\beta_{v})=\frac2{\ln2}\beta^2_{v}$.  In the third inequality, we assume the function of $\mathcal{G}_{v}$ is convex,  and in the last inequality, we assume $\mathcal{G}_{v}$  is  monotonically decreasing (which are held by the functions in the table in the main text).

\section{Supplementary note 7. Definitions and  bounds on genuine tripartite concurrence }\label{sect10}
Consider three-partite Mermin's inequality 
\begin{eqnarray*}
\beta_{\rm M}\equiv (A_{0}B_{1}+A_{1}B_{0})C_{0}+(A_{0}B_{0}-A_{1}B_{1})C_{1}\leq 2
\end{eqnarray*}
Here, we first recall the concurrence for genuine tripartite entanglement~\cite{PhysRevA.83.062325} $E^{(3)}_{\rm C}(\psi)$ for a pure state $|\psi \rangle$ as 
\begin{eqnarray}
E^{(3)}_{\rm C}(\psi)&\equiv& \sqrt{2-2\max_{i= A, B, C} \{ {\rm Tr} (\phi_{i}^2)\}}, 
\end{eqnarray}
where $\phi_{A}={\rm Tr}_{BC}(|\psi \rangle \langle \psi |)$ and the same definitions  for $\phi_{B, C}$.   For a mixed state $\rho$, genuine tripartite concurrence is defined as  
\begin{eqnarray}
  E^{(3)}_{\rm C}(\rho)\equiv \min_{\{r_{i}, \psi_{1}\}}\sum_{i}r_{i}E^{(3)}_{\rm C}(\psi_{i})
\end{eqnarray}
Noting that ${\rm Tr} (\phi^2_{A})= {\rm Tr}(|\psi\rangle \langle \psi|\varrho_{A|BC})$, where $\varrho_{A|BC}$  is the diagonal part of $|\psi \rangle \langle \psi|$ when written in Schmidt basis under a  partition  $A|BC$,  completely parallel to the consideration in the two-qubit case and using \ref{Jordan3}, we have 
\begin{eqnarray}
  E^{(3)}_{\rm C}(\rho)&\geq &\sum_{h}r_{h}E^{(3)}_{\rm C}(\rho_{h})\geq \sum_{h}r_{h}\sqrt{2} E_{\rm Tr}(\rho_{h}) \nonumber\\
  &\geq&  \sqrt{2}\tilde {\beta}^{(3)}_{v}\geq  \sqrt{2}{\beta }^{(3)}.
\end{eqnarray}
where $E^{(3)}_{\rm Tr}(\rho)$ specifies the minimal  trace-distance of concerned  $\rho$ to bi-partition separable states, namely, the ones assuming the decomposition $\varrho_{(21)}=\sum_{i}r_{i}\rho^{(i)}_{A}\otimes \rho^{(i)}_{BC} + \sum_{i}r'_{i}\rho^{(i)}_{B}\otimes \rho^{(i)}_{AC}+ \sum_{i}r''_{i}\rho_{C}^{(i)}\otimes \rho^{(i)}_{AB}$, and the relevant correlations formulate a set  specified as  $\mathcal{L}_{(21)}$,  and ${\beta }^{(3)}\equiv  \frac{\beta_{\rm M}-\beta(\mathcal{L}_{(21)})}{\beta_{\max}-\beta_{\min}}$,        $\tilde {\beta}^{(3)}_{v}\equiv  \frac{\beta_{\rm  M}-f_{(21)}(\beta_{\rm M})}{\beta_{\max}-\beta_{\min}}$, where  $\beta(\mathcal{L}_{(21)})$ is the maximum Bell value that can be obtained using state    $\varrho_{(21)}$ and $\tilde {\beta}^{(3)}_{v}$ is the optimized Bell's violation using strengthened bound. For Merming's inequality using qubit projection measurements,  we have    $f_{(21)}(\beta)=\frac{\beta}{2}+\sqrt{4-\frac{\beta^{2}}{4}}$  when $\beta> 2\sqrt{2}$ and   $f_{(21)}(\beta)=2\sqrt{2}$ when $\beta\leq  2\sqrt{2}$. Then it follows from the   concavity and monotonically decreasing  $f_{(21)}$  that  $\beta (\mathcal{L}'_{\bf P})\leq  f_{(21)}(\beta )$.   With the tightened  bound for fully separable state provided in Ref~\cite{sun2025}. we have bound for total entanglement  $E_{C}\geq  \sqrt{2}\tilde \beta_{v}$.

To give a brief idea about the  tightness  of our lower bounds,  an upper bound on the  minimal  entanglement that is sufficient   to  generate a Bell value  $\beta_{\rm M}\geq 2\sqrt{2} $  is given. Let us consider a simple  but non-trivial state $\rho_{\lambda}\equiv \lambda |\Psi \rangle \langle  \Psi|+ (1-\lambda )\varrho$, where state $|\Psi\rangle =1/{\sqrt{2}}(|000\rangle +|111\rangle ) $ is the  GHZ state that can achieve quantum bound $4$ and $\varrho$ is a partially entangled  state that can achieve the maximum  Bell value of  $2\sqrt{2}$.   Using this state and suitably chosen measurement settings, one can obtain a Bell quantity value as 
$$\beta_{\rm M} (\mathbf{P}_{\rho_{\lambda}})= 4\lambda +2\sqrt{2}(1-\lambda )=2\sqrt{2}+(4-2\sqrt{2})\lambda.$$
An upper bound of $E^{(3)}_{\rm C} (\rho_{\lambda })$ can be given by considering   a decomposition that may be not optimal,  $ \{\lambda, |\Psi\rangle , r_{i}, |\phi_{i}\rangle \}$,  where $\phi_{i}$ are bipartition-separable state. Then, an upper bound is given as   
 $E^{(3)}_{\rm C}(\rho_{\lambda})\leq \lambda E^{(3)}_{\rm C}(\Psi)+\sum_{i}r_{i}E^{(3)}_{\rm C}(\phi_{i})=\lambda $. 
Then we have 
$$E^{(3)}_{\rm Tr}(\rho_{\lambda})\leq \frac{\beta-2\sqrt{2}}{4-2\sqrt{2}}.$$
Then we argue that $ (\beta_{\rm M}-2\sqrt{2})/ (4-2\sqrt{2})$  is the maximum entanglement  that  a Bell value $\beta_{\rm M}$ can be verified.

\end{document}